\documentclass[preprint, nobibnotes, aps, superscriptaddress]{revtex4-1}
\usepackage{amsmath}
\usepackage{amssymb}
\usepackage{hyperref}
\usepackage{physics}
\usepackage{graphicx}
\usepackage{color}

\begin{document}

\title{Generalising the Yagi--Uda Antenna: Designing Disordered Metamaterials to Manipulate Antenna Radiation}

\author{J. R. Capers}
\email{j.capers@exeter.ac.uk}
\affiliation{Department of Physics and Astronomy, University of Exeter, Stocker Road, Exeter, EX4 4QL, United Kingdom}

\author{L. D. Stanfield}
\email{lds211@exeter.ac.uk}
\affiliation{Department of Physics and Astronomy, University of Exeter, Stocker Road, Exeter, EX4 4QL, United Kingdom}

\author{J. R. Sambles}
\affiliation{Department of Physics and Astronomy, University of Exeter, Stocker Road, Exeter, EX4 4QL, United Kingdom}

\author{S. J. Boyes}
\affiliation{DSTL Porton Down, Salisbury, Wiltshire, SP4 0JQ, United Kingdom}

\author{A. W. Powell}
\affiliation{Department of Physics and Astronomy, University of Exeter, Stocker Road, Exeter, EX4 4QL, United Kingdom}

\author{A. P. Hibbins}
\affiliation{Department of Physics and Astronomy, University of Exeter, Stocker Road, Exeter, EX4 4QL, United Kingdom}

\author{S. A. R Horsley}
\affiliation{Department of Physics and Astronomy, University of Exeter, Stocker Road, Exeter, EX4 4QL, United Kingdom}

\date{\today}

\begin{abstract}
    Next generation microwave communications systems face several challenges, particularly from congested communications frequencies and complex propagation environments.
    Taking inspiration from the Yagi--Uda antenna, we present, and experimentally test, a framework based on the coupled dipole approximation for designing structures composed of a single simple emitter with a passive disordered scattering structure of rods that is optimised to provide a desired radiation pattern.
    Our numerical method provides an efficient way to model, and then design and test, otherwise inaccessibly large scattering systems.
\end{abstract} 

\maketitle

\section{Introduction}

In recent years metamaterials, man--made materials structured at the sub--wavelength scale, have attracted much interest due to their versatile wave--shaping capabilities.
From invisibility cloaks \cite{Leonhardt2006, Pendry2006} to perfect lenses \cite{Kaina2015}, metamaterials offer novel ways to control and to shape the propagation of electromagnetic, acoustic and elastic waves \cite{Kadic2019}.
While early examples of metamaterials consist of periodically patterned metal \cite{Munk2000} or dielectric \cite{Joannopoulos2008}, more recently disorder has been exploited \cite{Cao2022} to achieve a range of interesting wave effects such as anti--reflection coatings \cite{Horodynski2022}, energy storage \cite{Hougne2021}, perfect absorption \cite{Pichler2019}, analogue computing \cite{Sol2022} and imaging \cite{Mosk2012}.

Disordered materials have also been employed to manipulate the radiation from sub--wavelength emitters \cite{Brule2022, Wiecha2018a, Wiecha2019a}.
In this context, the role of disorder is to provide many more design degrees of freedom to achieve the desired behaviour.
Engineering the radiation from small sources has many key applications for the next generation of microwave communications.
Current networks face many challenges \cite{Akyildiz2020}  particularly from congested communications frequencies and complex propagation environments.
Rather than solving this complex propagation problem using a combination of a large numbers of antenna, one can instead try to move the functional complexity into a metamaterial layer that can reshape the radiation from a given single source in a desired way  and which may be easily fabricated and modified to change functionality.
A key question is, therefore, how to design such a metamaterial layer to structure in a specified way the radiation pattern of an emitter.

Several methods for designing passive scattering structures to manipulate the radiation from an emitter have been developed over the last decade.
Utilising elegantly simple phase--based arguments, structures have been designed \cite{Mignuzzi2019} and experimentally realised \cite{Stanfield2023} that enhance the radiation efficiency of an emitter by factors in the thousands.
Methods based on perturbation theory \cite{Bennett2022, Bennett2021} and shape optimisation \cite{Matuszak2022} have been utilised in quantum electrodynamics to control energy transfer and enhance coherence.  
Genetic algorithms have also been employed \cite{Brule2022, Wiecha2018a, Wiecha2019a} to distribute scattering elements to produce antenna with specific radiation properties.
In the context of antenna engineering, disordered phased arrays have been designed using genetic algorithms \cite{Roy2010} and sparse optimisation \cite{Silverstein2023, Nai2010, Yang2021}.

\begin{figure}[h!]
    \centering
    \includegraphics[width=\linewidth]{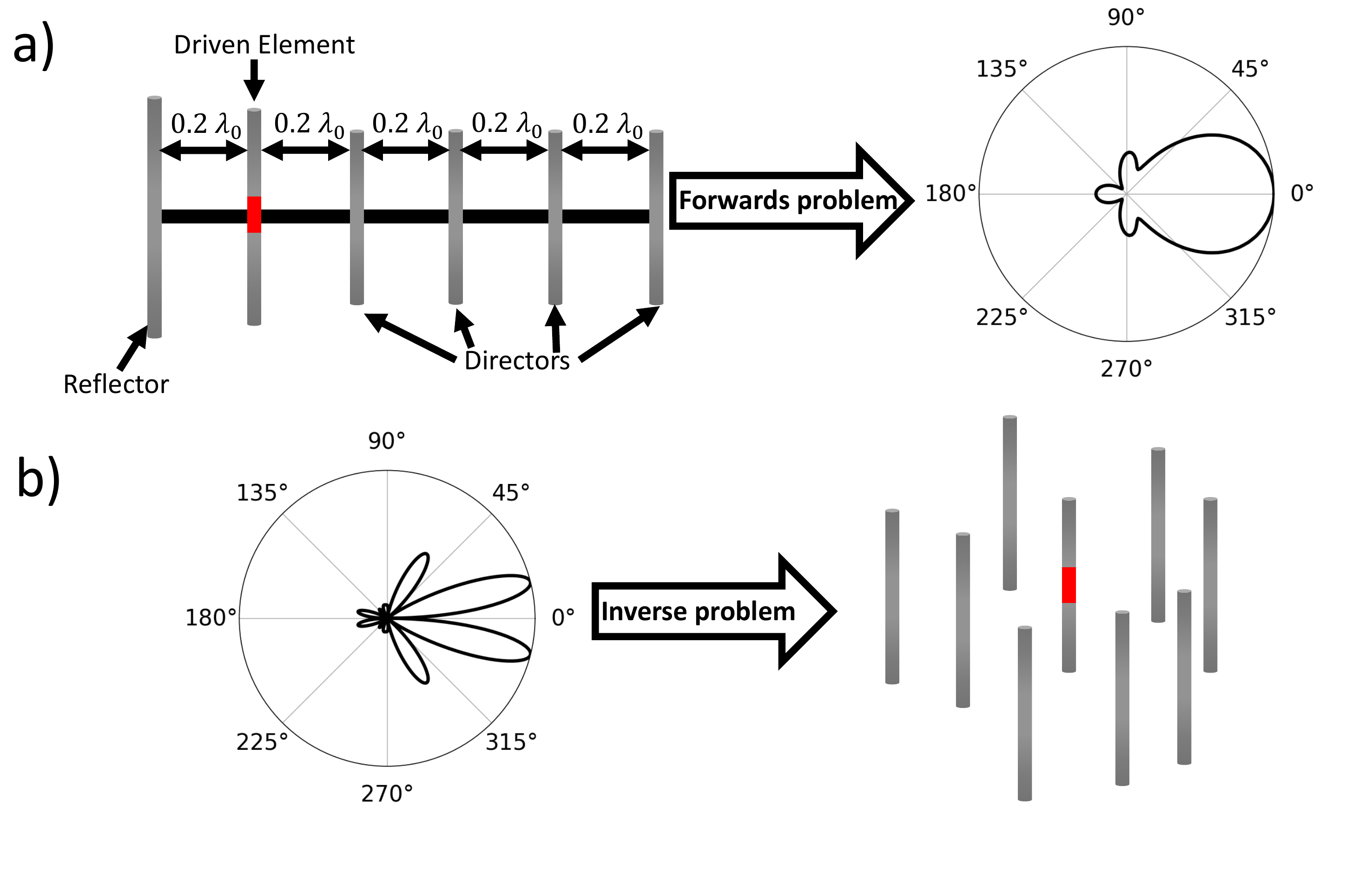}
    \caption{Schematic of the problem to be solved.
    A conventional Yagi--Uda antenna shown in a), is composed of a single driven element and many passive scattering elements, that, through constructive interference produce a directive beam.  
    We instead ask b) how an array of scattering rods should be distributed around an emitter to achieve a particular radiation pattern.}
    \label{fig:schematic}
\end{figure}

In this work, we consider designing disordered arrangements of scatterers (metal rods) around a single emitter to engineer the far--field radiation pattern in a particular plane.
One way to illustratively understand this is as a generalised case of the Yagi--Uda antenna \cite{Yagi1926}.
Composed of a single driven element, a reflector behind it along with several `director' elements supported by a dielectric boom, the Yagi--Uda antenna generates a highly directive beam in a single direction.
The behaviour of a conventional Yagi--Uda antenna is shown in Figure \ref{fig:schematic} a), where the parameters of the antenna have been taken from \cite{yagi_comsol}.
Rod lengths are chosen to correspond to the fundamental dipole (half--wavelength) resonance, while separations are chosen to maximise constructive interference in the prefered radiation direction.
Despite being almost 100 years old, the Yagi--Uda antenna continues to attract interest, providing a source of inspiration to the optics community \cite{Novotny2011, Kullock2020, Maksymov2012} as well as to antenna engineers \cite{Hao2021, Monti2016, Tehrani2015, Yang2018}.
Simple to fabricate, easy to design due to the relatively few tunable parameters and lightweight, the Yagi--Uda antenna is the obvious choice for line--of--sight communications.
Typically, one knows the geometry of the system i.e. the properties of the emitter and the positions and lengths of the rods, then calculates the radiation pattern produced.
This `forwards problem' is shown in Figure \ref{fig:schematic} a).
Here, instead, we formulate a solution to the `inverse problem', shown in Figure \ref{fig:schematic} b).
Given that we know the radiation pattern we would like, we develop an inverse design framework to optimise the position of the scattering rods to achieve this.
In this paper we design and experimentally realise several different structures with particular radiation patterns in the plane of the rods.
The key benefit of our approach is that our semi--analytic framework enables the design of large disordered systems that would otherwise be numerically intractable.

\section{Theory, Modelling and Inverse Design}

\begin{figure}[h!]
    \centering
    \includegraphics[width=\linewidth]{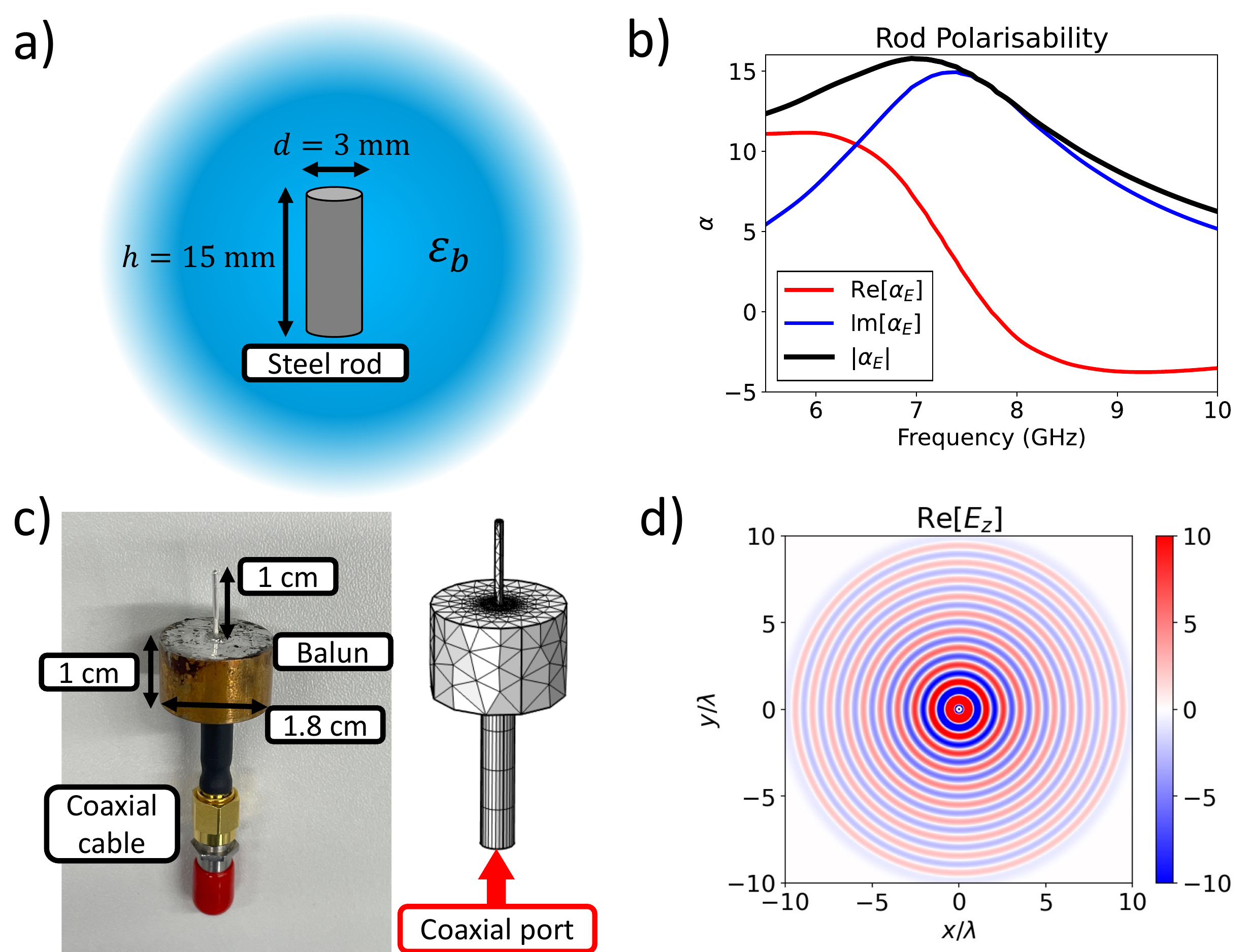}
    \caption{Finite element modelling of the scatterers and emitter.
    A schematic of the a) finite element model used to calculate the polarisability of the rods, with the resulting polarisability shown in b).
    Photograph and finite element representation of the source c), along with the d) calculated radiation field.}
    \label{fig:building_blocks}
\end{figure}
Before attempting to design the structures, we will develop an efficient way to model them so that all of the fields of the system can be extracted and analysed efficiently.
Modelling a system comprised of a large number of scattering elements around an emitter is generally very challenging.
In finite element or finite difference full--wave solvers the length scale separation between the sub--wavelength scattering elements and the tens of wavelengths that make up the whole structure can result in intractably large meshes.
To avoid this, we utilise the coupled dipole approximation \cite{Foldy1945, Purcell1973}.
Assuming that each of the passive scatterers can be described as a point dipole located at position $\boldsymbol{r}_n$, scattering from these introduces source terms in Maxwell's equations related to the polarisation and magnetisation densities
\begin{align}
    \boldsymbol{P} &= \sum_n \boldsymbol{\alpha}_E \boldsymbol{E} (\boldsymbol{r}_n) \delta (\boldsymbol{r}-\boldsymbol{r}_n)\\  
    \boldsymbol{M} &= \sum_n \boldsymbol{\alpha}_H \boldsymbol{H} (\boldsymbol{r}_n) \delta (\boldsymbol{r}-\boldsymbol{r}_n),
    \label{eq:PM}
\end{align}
where $\boldsymbol{\alpha}_E$ is the electric polarisability of the scatterer, $\boldsymbol{\alpha}_H$ is the magnetic polarisability.
These tensors relate the applied electric and magnetic fields, $\boldsymbol{E}$ and $\boldsymbol{H}$ to the polarisation and magnetisation densities $\boldsymbol{P}$ and $\boldsymbol{M}$.
More complicated scattering elements can be described in this framework if one includes additional multipolar terms in the electric and magnetic polarisabilities \cite{Evlyukhin2011, Evlyukhin2013}.
In this work, we restrict ourselves to the case where all of the scatterers can be described as dipoles.
With these source terms, the solution to Maxwell's equations can be written as 
\begin{equation}
    \phi (\boldsymbol{r}) = \phi_{\rm inc} (\boldsymbol{r}) + \sum_{n=0}^N \mathbb{G}(\boldsymbol{r}, \boldsymbol{r}_n) \phi (\boldsymbol{r}_n) ,
    \label{eq:waveSln}
\end{equation}
where we have adopted the compact notation of \cite{Capers2022}, that
\begin{align}
    \phi (\boldsymbol{r}) &= 
    \begin{pmatrix}
    \boldsymbol{E} (\boldsymbol{r}) \\
    \eta_0 \boldsymbol{H} (\boldsymbol{r})
    \end{pmatrix} \\
    \mathbb{G}(\boldsymbol{r}, \boldsymbol{r'}) &= 
    \begin{pmatrix}
        \xi^2 \boldsymbol{G}(\boldsymbol{r}, \boldsymbol{r'}) \boldsymbol{\alpha}_E & i \xi \boldsymbol{G}_{EH}(\boldsymbol{r}, \boldsymbol{r'}) \boldsymbol{\alpha}_H \\ 
        -i\xi \boldsymbol{G}_{EH}(\boldsymbol{r}, \boldsymbol{r'}) \boldsymbol{\alpha}_E & \xi^2 \boldsymbol{G}(\boldsymbol{r}, \boldsymbol{r'}) \boldsymbol{\alpha}_H
    \end{pmatrix} ,
\end{align}
where 
\begin{equation}
    \boldsymbol{G} (\boldsymbol{r}, \boldsymbol{r'}) = \left[ \boldsymbol{1} + \frac{1}{\xi^2} \nabla \otimes \nabla \right] \frac{e^{i\xi|\boldsymbol{r} - \boldsymbol{r'}|}}{4 \pi |\boldsymbol{r} - \boldsymbol{r'}|}
\end{equation}
is the Dyadic Green's function \cite{Schwinger1950}, $\xi = ka$ is a dimensionless wave--number with $a$ being a characteristic length scale of the problem, $\eta_0$ is the impedance of free space and $\boldsymbol{G}_{EH} = \nabla \times \boldsymbol{G}(\boldsymbol{r}, \boldsymbol{r'})$.
It should be noted that the fields applied to each scatterer $\phi (\boldsymbol{r}_n)$ must be calculated by imposing self--consistency \cite{Capers2021, Capers2022}, resulting in a matrix of size $6N \times 6N$ that must be inverted.
This is the main computational burden of the coupled dipole method, but ensures that all mutual coupling effects are fully accounted for.

Inverse design of the system is then facilitated according to the method presented by Capers et al. \cite{Capers2021}.
Starting from an initial distribution of scattering elements, the aim is to derive an analytic expression to find how to iteratively move the scatterers in order to increase a chosen figure of merit.
To do this, we ask how moving a scatterer by a small amount changes the fields of the system.
Perturbatively expanding the position of the scatterers
\begin{equation}
    \delta (\boldsymbol{r} - \boldsymbol{r}_n - \delta \boldsymbol{r}_n) = \delta (\boldsymbol{r} - \boldsymbol{r}_n) - (\delta \boldsymbol{r}_n \cdot \nabla ) \delta (\boldsymbol{r} - \boldsymbol{r}_n) + \ldots ,
    \label{eq:fieldVar}
\end{equation}
one can find \cite{Capers2021, Capers2022} that the change produced in the field is
\begin{equation}
    \delta \phi (\boldsymbol{r}) = \mathbb{G}(\boldsymbol{r}, \boldsymbol{r}_n) \nabla \phi (\boldsymbol{r}_n) \cdot \delta \boldsymbol{r}_n .
    \label{eq:field_var}
\end{equation}
As our aim is to shape the far--field radiation pattern of the system, we define as our figure of merit the overlap integral in the far--field in the plane of the scatterers
\begin{equation}
    \mathcal{F} = \frac{1}{\mathcal{N}} \int_0^{2\pi} d\phi |\boldsymbol{E}(\phi)| |\boldsymbol{E}_{\rm target}(\phi)| ,
    \label{eq:ol_int}
\end{equation}
where $\mathcal{N}$ is a normalisation factor that depends upon both $\boldsymbol{E}$ and $\boldsymbol{E}_{\rm target}$.
This figure of merit ranges from 0 to 1, when the radiation pattern is exactly the same as the target radiation pattern.
This can be expanded under small changes of the field \cite{Capers2022} to find an analytic expression for the gradient of the figure of merit with respect to the positions of the scatterers.
The expression in given in the Supplementary Material.
With this gradient found, elementary gradient descent optimisation
\begin{equation}
    \boldsymbol{r}_n^{i+1} = \boldsymbol{r}_n^i + \gamma \nabla_{\boldsymbol{r}_n} \mathcal{F}
\end{equation}
with $\gamma$ being the step size, can be used to iteratively update the positions of the scatterers to increase the figure of merit.

In order to use this framework to describe our experimental system, certain properties of the scatterers and the emitter must be extracted numerically.
For scatterers, we use metal rods of length 15 mm and diameter 3 mm.
These parameters were chosen so that the dipole resonance of the rods would be at $\sim 10$ GHz, although any sub--wavelength resonator could be utilised.
The polarisability tensors, $\boldsymbol{\alpha}_E$ and $\boldsymbol{\alpha}_H$ can be found using full--wave simulations following the prescription of Yazdi et al. \cite{Yazdi2016, Yazdi2016a}.
Exciting the rod with plane waves from different directions, the induced charges and currents can be integrated to find the electric and magnetic dipole moments, which can be related to columns of the polarisability tensor.
A schematic of the finite element model is shown in Figure \ref{fig:building_blocks} a), with the resulting $z$ component of the polarisability as a function of frequency shown in Figure \ref{fig:building_blocks} b).
All other components of $\boldsymbol{\alpha}_E$ and $\boldsymbol{\alpha}_H$ are negligible.
Electing to work at 7 GHz, the dipole resonance of the rods, the polarisability is
\begin{equation}
    \boldsymbol{\alpha}_E = 
    \begin{pmatrix}
        0 & 0 & 0 \\
        0 & 0 & 0 \\
        0 & 0 & 1
    \end{pmatrix}
    (6.91 + i14.17) .
\end{equation}
For the scatterers to be lossy, the polarisability tensor must obey the inequality ${\rm Im}[\boldsymbol{\alpha}_E] \geq \boldsymbol{1} k^3 / (6 \pi \varepsilon)$ \cite{Belov2003}, where $\varepsilon$ is the permittivity of the background medium containing the scatterer.
The other key component of the couple dipole approximation Eq. \ref{eq:waveSln} that we require is the field due to the source emitter.
Our choice of emitter, a sleeve antenna, is shown in Figure \ref{fig:building_blocks} c).
The fields of the emitter on its own are calculated using COMSOL Multiphysics \cite{comsol}, shown in Figure \ref{fig:building_blocks} d), then exported and interpolated to provide a function for the source fields that is fast to evaluate.
As the scatterers have only a single non--zero component of their polarisability tensors, only the $z$ component of the electric field is relevant.
Correctly representing the source field is key to using this method to correctly design wave--shaping devices.
As such, a complete numerical and experimental characterisation of the emitter is included in Supplementary Material.

With the polarisability of the scatterers and the emitter characterised, the coupled dipole framework can be used to both model and design a system with a specific radiation pattern.
An investigation of how well the coupled dipole framework describes the system is included in the Supplementary Material.
The key benefit of our approach is that the numerical complexity of the model depends upon the number of scatterers $N$, involving the inversion of a matrix of the size $6N \times 6N$.
Conversely, most full--wave solvers must solve linear systems that depend upon the size of the computational domain.
For the structures we consider here, this could be a volume of $\sim 1000 \lambda^3$, resulting in hundreds of thousands of mesh elements.
Indeed, describing large systems of aperiodic sub--wavelength scatterers in this way currently requires specialist computers, while the coupled dipole framework can run on a modest laptop.
Consequently, our approach enables the design of otherwise inaccessible structures.

\section{Experimental Results}

Utilising the modelling and design framework based on the coupled dipole approximation, we have designed several devices that illustrate both the capabilities and limitations of our novel approach.
Keeping the scatterers in the plane of the emitter, for simplicity, the design approach outlined in the previous section was employed to design structures of 100 metal rods, driven by a central emitter, that have specified radiation patterns.
Once designed and constructed, the structures were experimentally characterised using an Anritsu  MS46122B Vector Network Analyser (VNA), and a rotational table controlled by a Thorlabs APT Precision Motion Controller. 
To determine the far-field radiation pattern, the structure underwent a full 360$^{\circ}$ rotation in 1$^{\circ}$ increments about the z-axis, parallel to the axis of the source and the rod scatterers, in an anechoic chamber, with response measured by a Narda Standard Gain Horn (Model 642) for frequency range 4.0 GHz to 10.0 GHz in 0.05 GHz increments.
A photograph of the experimental setup is given in the Supplementary Material.

\begin{figure}[h!]
    \centering
    \includegraphics[width=\linewidth]{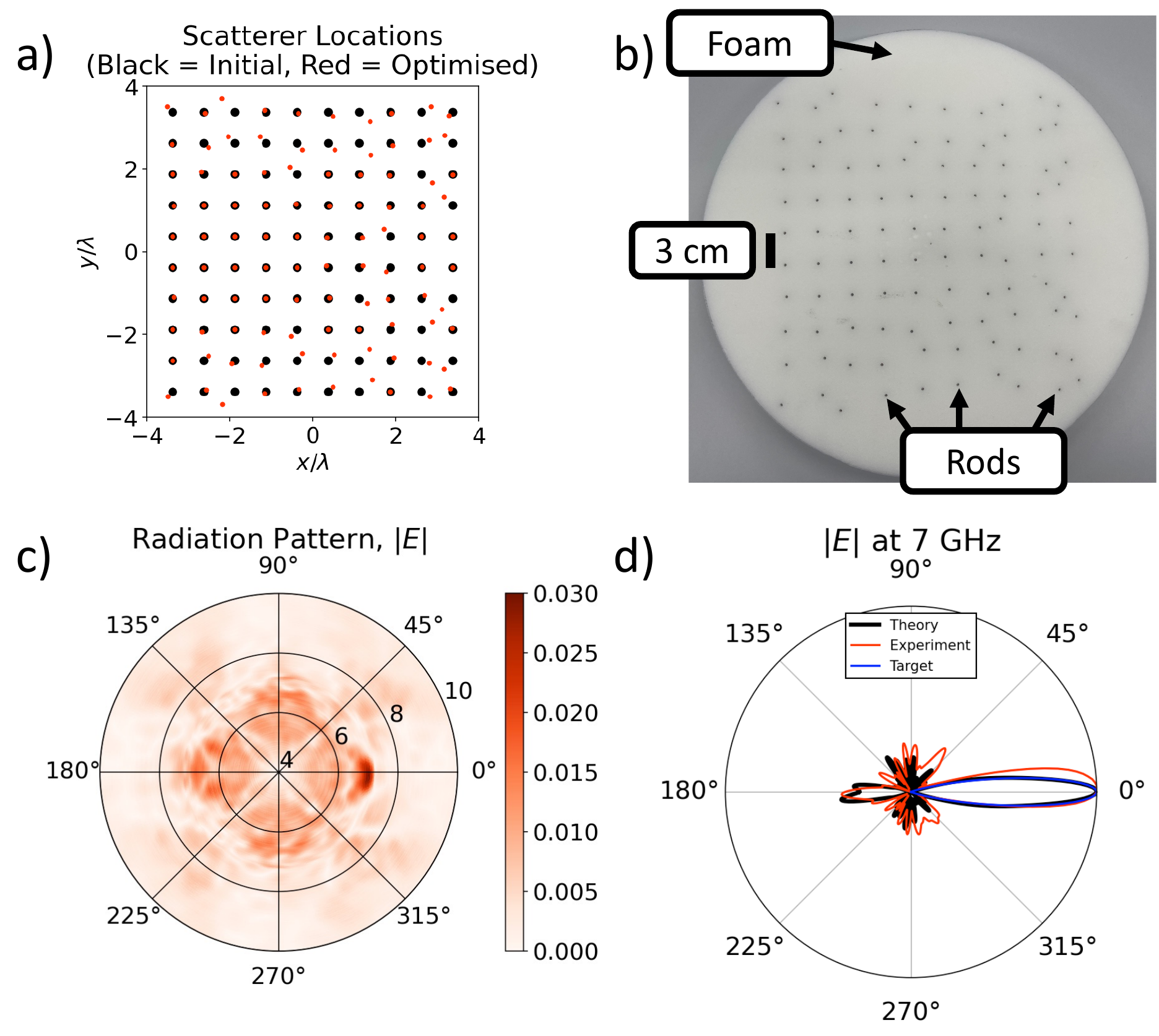}
    \caption{Designing a 100 scatterer structure giving a beam in a single direction of width 10$^\circ$.
    The initial and final positions of the rods are shown in a), where the emitter is located at the origin. 
    The experimental sample is shown in b), with the radiation pattern data given in c), where the radial axis is frequency in GHz.
    Taking a cut of this data at 7 GHz, the theory, experimental and desired radiation patterns are presented in d).}
    \label{fig:one_beam}
\end{figure}

The first example we present is shown in Figure \ref{fig:one_beam}.
Our target radiation pattern is a unidirectional beam at 0$^\circ$ with a beam width of 10$^\circ$.
We decide to undertake our optimisation based on an initial sqaure array of 100 rods with periodicity $0.75 \lambda \sim 3$ cm, shown in Figure \ref{fig:one_beam} a).
Optimising the overlap integral Eq. \ref{eq:ol_int}, the scatterers are re--distributed.
A low index disc of  Evonik Industries Roacelle IG foam, of permittivity $1.05$ and loss tangent 0.0017 at 10 GHz, is used to support the rods. which sit snuggly within the disc of thickness equal to the height of the rods.
The constructed device is shown in Figure \ref{fig:one_beam} b), with the measured radiation pattern shown in panel c).
A clear peak can be seen at the design frequency of 7 GHz.
Plotting the radiation pattern on a linear scale in Figure \ref{fig:one_beam} b), a beam at the correct location with a width close to the desired width is observed.

\begin{figure}[h!]
    \centering
    \includegraphics[width=\linewidth]{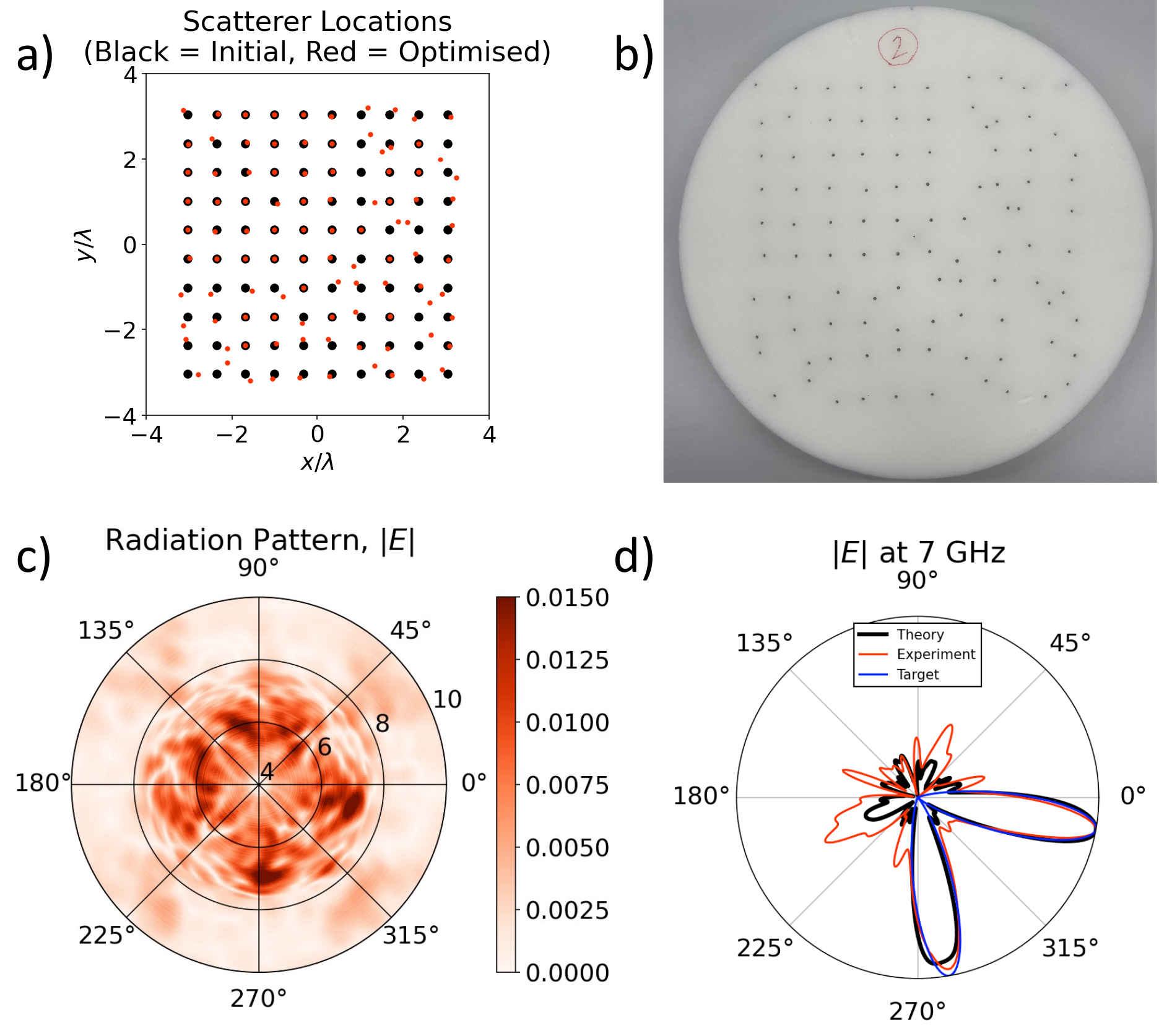}
    \caption{Designing a device that beams into 350$^\circ$ and 260$^\circ$, with beam widths of 15$^\circ$.
    Initial and final rod locations are shown in a), with the fabricated sample in b).
    The measured radiation pattern is shown in c), with the radial axis being frequency in GHz.
    A cut of the far field taken at 7 GHz is compared with theoretical predictions in d).}
    \label{fig:two_beam}
\end{figure}

In our next example, we design a device that has two beams of defined width at particular angles in the plane of the array.
We choose to place one beam at 350$^\circ$ and another at 260$^\circ$, both of width 15$^\circ$.
The associated device is shown in Figure \ref{fig:two_beam}.
Locations of the scatterers before and after the optimisation are shown in Figure \ref{fig:two_beam} a), with the fabricated structure shown in Figure \ref{fig:two_beam} b).
In the measured radiation pattern, Figure \ref{fig:two_beam} c) and d), one can observe the expected features.
Lobes at 350$^\circ$ and 260$^\circ$ are present and have the correct width.

\begin{figure}[h!]
    \centering
    \includegraphics[width=\linewidth]{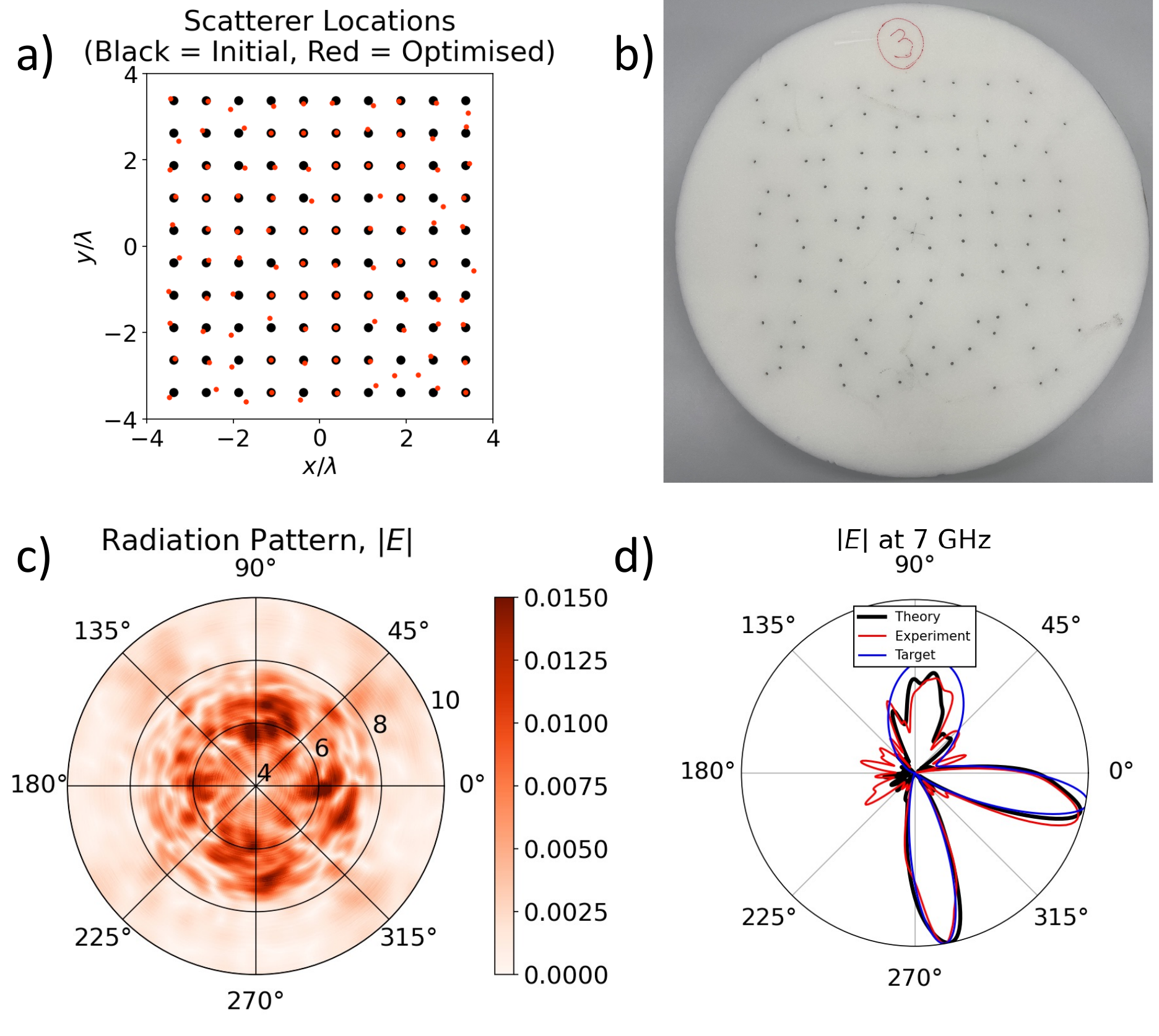}
    \caption{Designing a device with two beams of maximum amplitude and width 15$^\circ$ at 260$^\circ$ and 350$^\circ$, plus a beam of half amplitude at 80$^\circ$ with width 45$^\circ$.
    Initial and final rod locations are shown in a), with the experimental sample shown in b).
    Measured radiation pattern is shown in c), where the radial axis is frequency in GHz, with a comparison between experiment and theory shown in d).}
    \label{fig:three_beam}
\end{figure}

Finally, to demonstrate the flexibility of our method we design a device that has beams of different widths in three different directions, with also different relative far--field amplitudes.
Shown in Figure \ref{fig:three_beam}, we aimed to design a device that has two beams of width 15$^\circ$ at 350$^\circ$ and 260$^\circ$ at maximum amplitude and a beam at 80$^\circ$ of width 45$^\circ$ at half maximum amplitude.
The locations of the rods before and after the optimisation are shown in Figure \ref{fig:three_beam} a), with the device shown in Figure \ref{fig:three_beam} b).
The measured radiation pattern, Figure \ref{fig:three_beam} c), shows the maximum amplitude beams at 200$^\circ$ and 250$^\circ$, with the half amplitude beam at around 100$^\circ$.
Taking a slice at 7 GHz, Figure \ref{fig:three_beam} d), we see that the two full--amplitude beams occur at the correct positions and have the correct widths.
The half--amplitude beam is at the correct position with approximately correct width and amplitude. 

\begin{figure*}[h!]
    \centering
    \includegraphics[width=\linewidth]{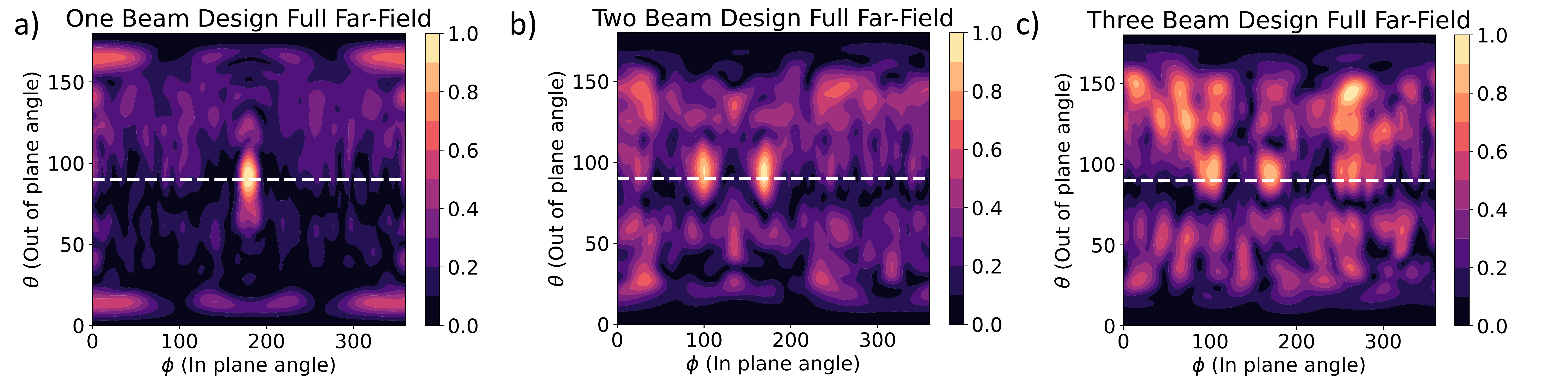}
    \caption{The full far--field spheres of the devices shown in a) Figure \ref{fig:one_beam}, b) Figure \ref{fig:two_beam} and c) Figure \ref{fig:three_beam}.
    White dashed line indicates the plane in which the radiation pattern has been optimised in.}
    \label{fig:full_ffs}
\end{figure*}

While we aim to shape the radiation in the plane of the emitters, for completeness the full far--field radiation pattern sphere for each device is shown in Figure \ref{fig:full_ffs}.
The white dashed line indicates the plane in which the radiation pattern has been structured.
One can observe that for both the one and two beam structures, Figure \ref{fig:full_ffs} a) and b) respectively, most of the radiation is localised to the plane of the emitters.
However, the radiation pattern of the three lobe structure Figure \ref{fig:full_ffs} c), shows significant out--of--plane radiation.
To avoid this unwanted radiation, one could define the figure of merit as an overlap integral over the full far--field.

\begin{figure*}[h!]
    \centering
    \includegraphics[width=\linewidth]{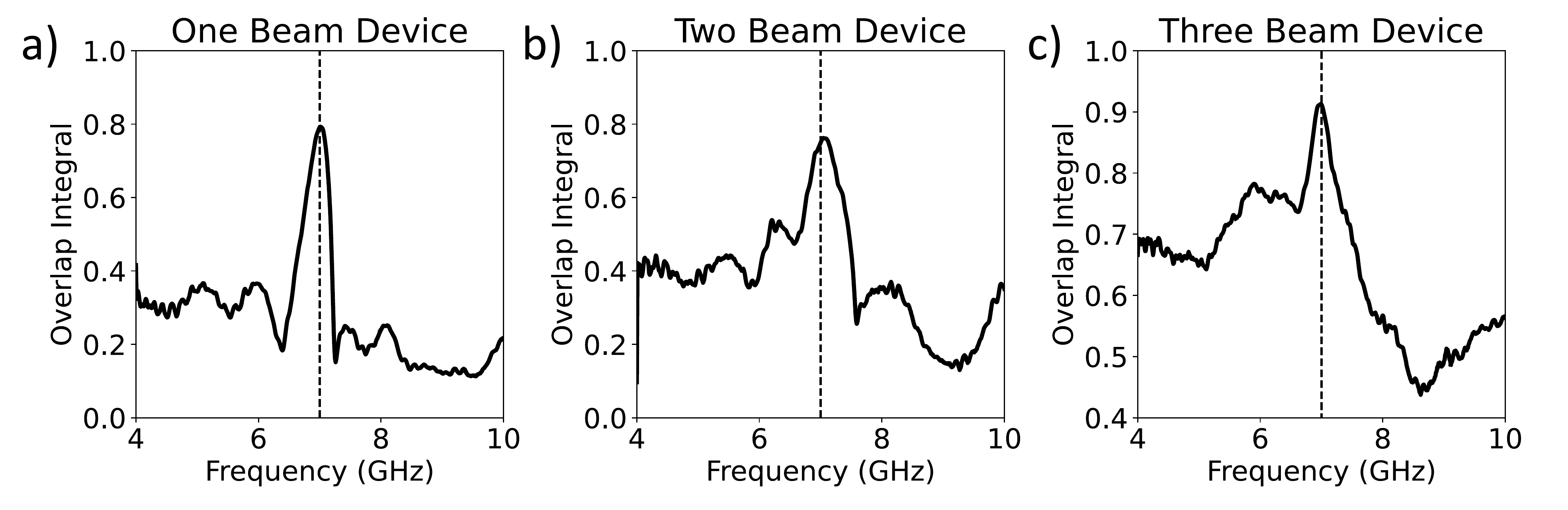}
    \caption{Frequency response of our devices, calculated by evaluating the overlap integral Eq. \ref{eq:ol_int}, with the measured radiation patterns.
    Dashed line indicates 7 GHz, the design frequency.}
    \label{fig:bandwidth}
\end{figure*}

We also consider the bandwidth of our devices.
This is calculated by evaluating the overlap integral Eq. \ref{eq:ol_int} between the target distribution and the measured radiation pattern as a function of frequency and is shown in Figure \ref{fig:bandwidth}.
All of the devices are rather narrow--band, or frequency selective, due to the high order multiple--scattering effects that lead to the desired behaviour.
Each device performs the best at around 7 GHz, as expected.
Measurements of the return loss $|S_{11}|$ of our devices are given in Supplementary Material.
For the one and two beam devices $|S_{11}|$ is comparable to that of the emitter in free space.
The three beam device, however, exhibits an increase in return loss from $|S_{11}| = 0.1$ to $|S_{11}| = 0.3$.

While these experimental results show clear proof--of--concept, experimental and theoretical results do not perfectly align.
Many steps in constructing both the experiments and the semi--analytic design framework introduce non--negligible sources of error.
Approximating the rods as dipoles is well justified, as the quadrupole moment of them has been calculated to be $\sim 5$ orders of magnitude lower than the dipole moment.
However, treating them as point dipoles, rather than distributed currents, introduces deviations in the radiation pattern due to near--field effects.
Provided that the rods are separated by a distance larger than 5 times the radius, as is the case in all of the designs we have presented, this induces a small but constant error in the far--field radiation pattern.
Next, the effect of small errors in the measurement of the permittivity of the foam was considered.
An over or under estimation of the permittivity by $\sim 1\%$ was found to have little effect, however once the error reached $\sim 5\%$ large differences in both the main and back lobes were observed.
Importantly, the back lobes become larger relative to the main lobes, as we observe in the experimental data.
Furthermore, errors in the estimation of the polarisability of the rods can also introduce additional back lobes.
A slightly different $\boldsymbol{\alpha}_E$ could be caused by small inconsistencies in the lengths of the rods.
We find that a $5\%$ difference in the length of the rods can change the polarisability enough to introduce the back lobes observed in the experimental data.
One way to remedy this could be to infer $\boldsymbol{\alpha}_E$ from experimentally measured radiation patterns.
The full data generated for these investigations is given in Supplementary Note 4.

\section{Conclusions \& Outlook}

    We have experimentally demonstrated a new framework for manipulating the radiation from simple sources.
    To meet the challenges of the next--generation of communications, antenna functionality can be moved into the design of a passive material instead of the antenna itself,  so that feeding remains simple but functionality can be complex.
    The paradigm we have presented here is both simple and versatile.
    Passive scattering elements and the emitter are characterised by full--wave simulations independently, then the coupled--dipole approximation is used to model the whole system.
    While this approach introduces some numerical errors, it enables the modelling of otherwise numerically inaccessible systems.
    Inverse design of the scattering systems has then been performed using gradient descent, with gradients that can be efficiently evaluated as we have derived their analytic form.
    Several experimental results are presented that illustrate both the utility and limitations of our method.

    Future work might focus on the design of switchable or time--varying structures, so that antenna functionality can be tuned in real time.
    While our results have focused on a single polarisation, more exotic polarisation manipulation could be achieved with bianiostropic or chiral passive scatterers.
    Broadening bandwidth by including multiple species of scatterers supporting resonances at different spectral locations, or of different multipolar nature, would also be of great utility in the context of antennas.

\section*{Acknowledgements}

    J.R.C. would like to thank Dr. D. A. Patient and Dr. I. R. Hooper for many useful discussions.
    
    We acknowledge financial support from the Engineering and Physical Sciences Research Council (EPSRC) of the United Kingdom, via the EPSRC Centre for Doctoral Training in Metamaterials (Grant No. EP/L015331/1).  
    J.R.C also wishes to acknowledge financial support from Defence Science Technology Laboratory (DSTL).
    L. D. S. wishes to acknowledge financial support from Leonardo Ltd UK via the iCASE studentship grant EP/R511924/1.
    S.A.R.H acknowledges financial support from the Royal Society (URF\textbackslash R\textbackslash 211033).

\section*{Data Availability Statement}
    
    All data and code created during this research are openly available from the corresponding authors, upon reasonable request.


\end{document}